\documentclass[11pt]{article}%
\usepackage{makeidx, siunitx, ragged2e}
\usepackage{booktabs,lscape,color, float}
\usepackage{subcaption} 
\usepackage{amsfonts,bm}
\usepackage{amsmath}
\usepackage{amssymb}
\usepackage{graphicx}
\usepackage{rotating}
\usepackage{multirow}%
\usepackage{geometry,setspace}
\usepackage{natbib}
\usepackage{mathtools}
\usepackage{url} 
\usepackage{lipsum}
\begin{document}

\title{Choosing the Right Return Distribution and the Excess Volatility Puzzle}
\author{
Abootaleb Shirvani\thanks{Texas Tech University, Department of Mathematics
\& Statistics, Lubbock TX 79409-1042, U.S.A., abootaleb.shirvani@ttu.edu.}
  \and
Frank J. Fabozzi\thanks{EDHEC Business School, U.S.A., frank.fabozzi@edhec.edu (Corresponding
	Author).}
}
\maketitle

\begin{abstract}
Proponents of behavioral finance have identified several “puzzles”  in the market that are inconsistent with rational finance theory. One such puzzle is the “excess volatility puzzle”. Changes in equity prices are too large given changes in the fundamentals that are expected to change equity prices. In this paper, we offer a resolution to the excess volatility puzzle within the context of rational finance. We empirically show that market inefficiency attributable to the volatility of excess return across time, is caused by fitting an improper distribution to the historical returns. Our results indicate that the variation of gross excess returns is attributable to poorly fitting the tail of the return distribution and that the puzzle disappears by employing a more appropriate distribution for the return data. The new distribution that we introduce in this paper that better fits the historical return distribution of stocks explains the excess volatility in the market and thereby explains the volatility puzzle. Failing to estimate the historical returns using the proper distribution is only one possible explanation for the existence of the volatility puzzle. However, it offers statistical models within the rational finance framework which can be used without relying on behavioral finance assumptions when searching for an explanation for the volatility puzzle.
\end{abstract}


\newpage
\section{Introduction}\label{sec:intro}
Proponents of behavioral finance have attacked the models put forth by proponents of traditional finance which assumes that market participants should behavior rationally in making financial decisions. For this reason, traditional finance proponents are commonly referred to as proponents of rational finance. Those in the behavioral finance camp have pointed to the historical behavior of the stock market that is inconsistent with financial theory based on rational finance as evidence for the failure of rational finance. These empirical observations that are inconsistent with rational finance are referred to as finance “puzzles”. One of the most striking puzzles is the volatility puzzle.

The volatility puzzle has been used to describe two observations in the market: idiosyncratic volatility puzzle and  the excess volatility puzzle. With respect to the former, there are several empirical studies that have investigated the relationship between future return and risk.  These studies have used various proxies for risk such as the expected return from the capital asset pricing model and idiosyncratic risk.  According to financial theory, there should be a positive relationship between expected return and risk. Many empirical studies do indeed report a positive relationship. However, several studies, starting with \citet{Ang2006} and almost a decade later by \cite{Stambaugh:2015}, have found that contrary to what finance theory would predict, there is an inverse relationship, idiosyncratic risk and future return. This has been labeled the \textit{idiosyncratic volatility puzzle} or simply the \textit{volatility puzzle} and as the \textit{low volatility anomaly} by the asset management community. Several studies have used other measures of risk such as idiosyncratic skewness, co-skewness, maximum daily return, and liquidity risk to explain the idiosyncratic volatility puzzle.  The behavioral finance camp has offered a solution based on investor sentiment and the lottery effect.  

The volatility puzzle that we focus on in this paper is the \textit{excess volatility puzzle}, first identified by \cite{Shiller:1981} and \cite{LeRoy:1981}. Excess volatility in the equity market means that changes in equity prices are too large given changes in the fundamentals that are expected to move equity prices. As \cite{Shiller:2003} notes “The most significant market anomaly that efficient market theory fails to explain is excess volatility. The idea that stock prices change more than they rationally should is more troubling for efficient market theorists than any other anomaly, such as the January effect or the day-of-the-week effect.”  The behavioral finance explanation for the existence of excess volatility offered by Shiller and others is that the influence of investor behavior attributable to either psychological or sociological beliefs is more important than the pricing models proposed by rational finance.

Earlier studies have measured excess volatility by comparing the stock price volatility to volatility bounds obtained from stock prices derived from an assumed dividend discount model.   To measure the variation in excess returns, \cite{LeRoy:2016} defined a variation measure based on a composite variable, $z_t$, that depends on the stochastic discount factor, the growth rate of dividends, and the price-dividend ratio. In the LeRoy and Lansing model (LLM), the gross return on a stock and a bond are written as a function of the composite variables $z_t^{(S)}$ and $z_t^{(B)}$ respectively. They demonstrated that if the conditional variances of $z^{\left(S\right)}_{t+1}$  and  $z^{\left(B\right)}_{t+1}$ are constant across date-$t$ events, then excess returns on a stock are unpredictable. In that case, markets are efficient.  In their model, LeRoy and Lansing assume that the composite variables $z_t^{(S)}$ and $z_t^{(B)}$ are both conditionally log-normal.

In this paper, we show that if the composite variables are independent and identically log-normally distributed, then the conditional variances of $z^{\left(S\right)}_{t+1}$  and  $z^{\left(B\right)}_{t+1}$ are not constant across date-$t$ events. Therefore, the excess returns on a stock are predictable (i.e., markets are inefficient). If the distribution of financial risk-factors are indeed non-Gaussian distributed and exhibit tails that are heavier than the normal distribution, the LLM can explain the excess volatility puzzle. In this paper, we revisit the puzzle by extending the \cite{LeRoy:2016} approach to accommodate rare events. We assume two different distributions for the composite variables. 

First, we assume that  the distribution of the composite variables exhibits a log-normal inverse Gaussian (NIG) distribution.\footnote{See \cite{Barndorff:1997}.} We propose this distribution because the class of NIG distributions is a flexible system of distributions that includes fat-tailed and skewed distributions. Moreover, the normal distribution is a special case of NIG \citep[see][]{Barndorff:1997}. We investigate the excess volatility puzzle by fitting the NIG distribution to the LeRoy and Lansing's log-composite variables and evaluate conditional variances of $z^{\left(S\right)}_{t+1}$  and  $z^{\left(B\right)}_{t+1}$. 
We find that the estimates of the conditional variances of $z^{\left(S\right)}_{t+1}$  and  $z^{\left(B\right)}_{t+1}$ obtained from the NIG
fitted model are constant across date-$t$ events and the variation of the excess return is almost  equal to zero. Thus, the excess returns on stock are unpredictable, and therefore markets are efficient. 

Second, we generalize the NIG distribution by defining a new fat-tail distribution, the normal compound inverse Gaussian (NCIG), assuming that  the distribution of the log-composite variables is NCIG. 
By fitting the NCIG distribution, the variability of excess return is constant across date-$t$ event and the estimates of excess returns variation declines to zero. Thus, the excess volatility puzzle can be explained by fitting a more appropriate distribution for the excess gross return and thereby markets are efficient when the excess volatility puzzle is considered.

There are three sections that follow in this paper. In the next section, Section 2, we define the LLM model and derive a formula for the excess log-return of stocks and bonds by assuming a NIG distribution for the the composite variables. In Section 3 we describe our data set and fit three distributions normal, NIG and NCIG distributions to the data. We demonstrate that the variability in the conditional variances of $z^{\left(S\right)}_{t+1}$  and  $z^{\left(B\right)}_{t+1}$ produced by the models is partially attributable to the poor fit of the tail of the return distribution. We show that both the NIG and NCIG distributions are flexible enough to demonstrate the efficiency of markets in terms of the excess volatility puzzle. Section 4 concludes the paper.

\section{Volatility Puzzle}

Behavioral finance proponents assert that investors believe that the mean dividend growth rate is more variable than it actually is. Similarly, price-dividend ratios and returns might also be excessively volatile because, as behavioralists claim, investors extrapolate past returns too far into the future when forming expectations of future returns.\footnote{\cite{Shafir:1997}, and \cite{Ritter:2002} claim that the variation in $\frac{P}{D}$ ratios and returns are due to investors mixing real and nominal quantities when forecasting future cash flows. \cite{Barberis:2001} show that the degree of loss aversion depends on prior gains and losses.}

\cite{LeRoy:2016} offered a model that provides an explanation for the excess volatility puzzle.  By extending their framework, we offer resolutions to the excess volatility puzzle within the theory of rational finance to demonstrate how the excess volatility puzzle can be  resolved.

In the LLM, $R^{(S)}_{t+1}$, $t=0,1,\dots $ is the stock gross return\footnote{For a dividend-paying stock, the gross return on a stock is $R_{t+1}=\frac{S_{t+1}+d_{t+1}}{S_t}$ where $S_{t}$ is the ex-dividend stock price at $t$ and $d_{t+1}$ is the dividend
	received in period $t+1$.} in the period $(t,t+1]$, and it has the representation \footnote{See \cite{LeRoy:2016}.}

\begin{equation}
	\label{gross_return}
	R^{(S)}_{t+1}=\frac{p^{(S)}_{t+1}+d_{t+1\ }}{p^{(S)}_t}=\left(\frac{z^{(S)}_{t+1}}{{\mathbb{E}}_tz^{(S)}_{t+1}}\right)\left(\frac{1}{M_{t+1\ }}\ \right),
\end{equation}
where $p^{\left(S\right)}_t={\mathbb{E}}_t\left[\frac{M_{t+1\ }}{M_{t\ }}\left(p^{\left(S\right)}_{t+1}+d_{t+1\ }\right)\right]$   
is the ex-dividend stock price at $t$, ${\mathbb{E}}_tz^{(S)}_{t+1}=\mathbb{E}\left({z^{\left(S\right)}_{t+1}}/{{\mathcal{F}}^{\ }_{t\ }}\right)$, ${\mathcal{F}}^{}_{t}$ denotes the information set consisting of all linear functions of the past returns available until time $t$, $d_{t+1\ }$ is the dividend received in $(t,t+1]$, $M_{t\ }$ is the stochastic discount factor at $(t,t+1]$, and 
\begin{equation}
	\label{Stock_composite_variable}
	z^{(S)}_{t+1}= M_{t+1}\left(\frac{d_{t+1\ }}{d_{t}}\ \right)\left(1+\frac{p^{\left(S\right)}_{t+1}}{d_{t+1\ }}\right).
\end{equation} 
\citet[p.~4]{LeRoy:2016} introduced $z^{(S)}_{t+1}$ as a composite variable that depends on the stochastic discount factor, the growth of dividends and the price-dividend ratio.

Similarly, in the LLM, the gross bond return $R^{(B)}_{t+1}$ in the period $(t,t+1]$, has the representation
\begin{equation}
	\label{gross_bond_return}
	R^{(B)}_{t+1}=\frac{1+ qp^{(B)}_{t+1}}{p^{(B)}_t}=\left(\frac{z^{(B)}_{t+1}}{{\mathbb{E}}_tz^{(B)}_{t+1}}\right)\left(\frac{1}{M_{t+1\ }}\ \right),
\end{equation}
where $p^{\left(B\right)}_t={\mathbb{E}}_t\left[\left(\frac{M_{t+1\ }}{M_{t\ }}\right)\left(1+qp^{\left(B\right)}_{t+1}\right)\right]={\mathbb{E}}_tz^{(B)}_{t+1}$  is the price at $t$ of a default-free bond,   $q\in (0,1)\ $ is the coupon-rate at time $t$, and 
\begin{equation}
	\label{bond_compostie_variable}
	z^{(B)}_{t+1}= M_{t+1}\left(1+qp^{\left(B\right)}_{t+1}\right).
\end{equation}

\noindent Thus, the stock excess return by taking the logs and subtracting \eqref{gross_return} from \eqref{gross_bond_return} has the form 
\begin{equation}
	\label{stock_excess_return}
	lnR^{(S)}_{t+1}-lnR^{\left(B\right)}_{t+1}={ln z^{\left(S\right)}_{t+1}\ }-ln{\mathbb{E}}_tz^{\left(S\right)}_{t+1}-{ln z^{\left(B\right)}_{t+1}\ }+ln{\mathbb{E}}_tz^{\left(B\right)}_{t+1}.
\end{equation}
In \eqref{stock_excess_return}, ${ln z^{\left(S\right)}_{t+1}\ }-ln{\mathbb{E}}_tz^{\left(S\right)}_{t+1}$ and ${ln z^{\left(B\right)}_{t+1}\ }-ln{\mathbb{E}}_tz^{\left(B\right)}_{t+1}$ are the errors in forecasting ${ln z^{\left(S\right)}_{t+1}}$ and ${ln z^{\left(B\right)}_{t+1}}$ respectively.

Next, in the LLM, it is assumed that conditional on  ${\mathcal{F}}^{\ }_{t}$, the composite variables $z^{\left(S\right)}_{t+1/{\mathcal{F}}^{}_{t}}$  and  $z^{\left(B\right)}_{t+1/{\mathcal{F}}^{}_{t}}$ are log-normally distributed; in the other words, $ ln z^{\left(S\right)}_{t+1/{\mathcal{F}}^{}_{t}} \sim \mathcal{N}\left({\mu }^{(S)}_t,{{\sigma }^{(S)}_t}^2\right)$ and  $ln z^{\left(B\right)}_{t+1//{\mathcal{F}}^{\ }_{t\ }}\sim \mathcal{N}\left({\mu }^{(B)}_t,{{\sigma }^{(B)}_t}^2\right)$.\footnote{Here, $\sim$  stands for equal in distribution between two random variables or two stochastic processes.} This assumption for the distribution of the composite variables yields the following expression for the excess return:  
\begin{equation}
	\label{Normal_excess_return}
	lnR^{\left(S\right)}_{t+1/{\mathcal{F}}^{\ }_{t\ }}-lnR^{\left(B\right)}_{t+1/{\mathcal{F}}^{\ }_{t\ }}={ln z^{\left(S\right)}_{t+1/{\mathcal{F}}^{\ }_{t\ }}\ }-\left({\mu }^{\left(S\right)}_t+\frac{1}{2}{{\sigma }^{\left(S\right)}_t}^2\right)-{ln z^{\left(B\right)}_{t+1/{\mathcal{F}}^{\ }_{t\ }}\ }+\left({\mu }^{\left(B\right)}_t+\frac{1}{2}{{\sigma }^{\left(B\right)}_t}^2\right).
\end{equation}

By taking the mathematical expectation of both sides of \eqref{Normal_excess_return}, the excess gross return forecast is 

\begin{equation}
	\label{forecast_excess_gross_return} 
	\begin{array}{llll}
		{\mathbb{E}}_tlnR^{\left(S\right)}_{t+1}-{\mathbb{E}}_tlnR^{\left(B\right)}_{t+1}&=&
		{\mathbb{E}}_t{ln z^{\left(S\right)}_{t+1//{\mathcal{F}}^{\ }_{t\ }}\ }-\left({\mu }^{\left(S\right)}_t+\frac{1}{2}{{\sigma }^{\left(S\right)}_t}^2\right)\\
		&-&{\mathbb{E}}_t{ln z^{\left(B\right)}_{t+1//{\mathcal{F}}^{\ }_{t\ }}\ }+\left({\mu }^{\left(B\right)}_t+\frac{1}{2}{{\sigma }^{\left(B\right)}_t}^2\right)\\
		&=&-\frac{1}{2}{{\sigma }^{\left(S\right)}_t}^2+\frac{1}{2}{{\sigma }^{\left(B\right)}_t}^2. 
	\end{array}
\end{equation} 

\noindent Taking the unconditional variance in \eqref{forecast_excess_gross_return}, leads to

\begin{equation} 
	\label{predictable_variation_Normal} 
	\begin{array}{lll}
		Var\left({\mathbb{E}}_tlnR^{\left(S\right)}_{t+1}-{\mathbb{E}}_tlnR^{\left(B\right)}_{t+1}\right)&=&\frac{1}{4}Var\left({{\sigma }^{\left(S\right)}_t}^2-{{\sigma }^{\left(B\right)}_t}^2\right)\\ &=& 
		\frac{1}{4}Var\left({Var}_t\left({ln z^{\left(S\right)}_{t+1//{\mathcal{F}}^{\ }_{t\ }}\ }\right)-{Var}_t\left({ln z^{\left(B\right)}_{t+1//{\mathcal{F}}^{\ }_{t\ }}\ }\right)\right). 
	\end{array}
\end{equation}

Based on \eqref{predictable_variation_Normal}, \citet[p.~6]{LeRoy:2016} claim: 

\hfill\begin{minipage}{\textwidth-.5cm}
\textbf{LL-Efficiency Claim:} \textit{``The left-hand side of \eqref{predictable_variation_Normal} gives a measure of the  variation in excess returns. Eq. \eqref{predictable_variation_Normal} shows that if the conditional variances of } $z^{\left(S\right)}_{t+1/{\mathcal{F}}^{}_{t}}$  and  $z^{\left(B\right)}_{t+1/{\mathcal{F}}^{\ }_{t\ }}$ \textit{ are constant across date-$t$ events (although generally not equal to zero), then excess returns on stock are unpredictable. In that case markets are efficient. If, on the other hand, the conditional variances differ according to the event, then a strictly positive fraction of excess returns are forecastable, so markets are inefficient.''}\\
\end{minipage}

We show that because  the distribution of financial risk factors is non-Gaussian distributed and exhibits tails heavier than the Gaussian distribution, this can explain the efficiency of the market 
according to the \cite{LeRoy:2016} approach.

We now extend LLM, assuming that the composite variables, $z^{\left(S\right)}_{t+1/{\mathcal{F}}^{\ }_{t\ }}$  and  $z^{\left(B\right)}_{t+1/{\mathcal{F}}^{\ }_{t\ }}$ are log-NIG distributed. Recall that a random variable $X$  has a NIG distribution, denoted $X\sim NIG\left( \mu,\alpha,\beta,\delta\right) $, $  \mu \in \mathbb{R},\,\, \alpha\in \mathbb{R},\,\beta \in \mathbb{R},\,\delta \in \mathbb{R},\, \alpha^2>\beta^2$,
if its density is given by

\begin{equation}
	\label{NIG_dis}
	f_{X} \left( x\right) = \frac{\alpha \delta K_1 \left( \alpha \sqrt{ \delta^2 + \left( x-\mu \right) ^2 }\right) } { \pi \sqrt{\delta^2+\left( x-\mu\right) ^2 }} \exp {\left(  \delta \sqrt{ \alpha^2-\beta^2 }+\beta \left( x-\mu \right)\right) } , \,\,x \in \mathbb{R}.
\end{equation}

Then, $X$ has mean $E\left( X\right) =\mu +\frac{\delta \beta}{ \sqrt{\left( \alpha^2 -\beta^2 \right) }}$, variance $Var\left( X\right) =\frac{ \delta \alpha^2 }{\sqrt{\left( \alpha^2 -\beta^2 \right) ^{3} }}$ , skewness $\gamma\left( X\right) =\frac{3\beta}{\alpha \sqrt[4]{\delta^2 \left( \alpha^2-\beta^2\right) }}  $and excess kurtosis $\kappa\left( X\right) =\frac{3\left( 1+\frac{ 4\beta^2} {\alpha^2 }\right) }{\delta\sqrt{ \alpha^2-\beta^2 } }$. The moment-generating function of  $X$, $M_X\left(t\right)=\mathbb{E}e^{itX},\ t\in R,$ is given by

\begin{equation}
	\label{GrindEQ__12_} 
	M_X\left(t\right)={\mathrm{exp} \left\{\mu t+\delta \left(\sqrt{{\alpha }^2-{\beta }^2\ }-\sqrt{{\alpha }^2-{(\beta +t)}^2}\right)\right\}\ } 
\end{equation}

\noindent  Now, we assume that the composite variables $z^{\left(S\right)}_{t+1/{\mathcal{F}}^{\ }_{t\ }}$  and  $z^{\left(B\right)}_{t+1/{\mathcal{F}}^{\ }_{t\ }}$ are log-NIG distributed, denoted \[z^{\left(S\right)}_{t+1/{\mathcal{F}}^{\ }_{t\ }}\mathrm{\sim }lnNIG\left({\mu }^{\left(S\right)}_t,{\alpha }^{\left(S\right)}_t,{\beta }^{\left(S\right)}_t,{\delta }^{\left(S\right)}_t\right),\,\, \,\,\,  z^{\left(B\right)}_{t+1/{\mathcal{F}}^{\ }_{t\ }} \mathrm{\sim }lnNIG\left({\mu }^{\left(B\right)}_t,{\alpha }^{\left(B\right)}_t,{\beta }^{\left(B\right)}_t,{\delta }^{\left(B\right)}_t\right).\] 
Therefore, the excess gross return is 
\begin{align*} 
	\label{NIG_ excess_ gross_return}
	lnR^{\left(S\right)}_{t+1/{\mathcal{F}}^{\ }_{t\ }}-lnR^{\left(B\right)}_{t+1/{\mathcal{F}}^{\ }_{t\ }}={ln z^{\left(S\right)}_{t+1/{\mathcal{F}}^{\ }_{t\ }}\ }-\left({\mu }^{\left(S\right)}_t+{\delta }^{\left(S\right)}_t\left(\sqrt{{{\alpha }^{\left(S\right)}_t}^2-{{\beta }^{\left(S\right)}_t}^2\ }-\sqrt{{{\alpha }^{\left(S\right)}_t}^2-{\left({\beta }^{\left(S\right)}_t-1\right)}^2}\right)\right)\\
	-{ln z^{\left(B\right)}_{t+1/{\mathcal{F}}^{\ }_{t\ }}\ }+\left({\mu }^{\left(B\right)}_t+{\delta }^{\left(B\right)}_t\left(\sqrt{{{\alpha }^{\left(B\right)}_t}^2-{{\beta }^{\left(B\right)}_t}^2\ }-\sqrt{{{\alpha }^{\left(B\right)}_t}^2-{\left({\beta }^{\left(B\right)}_t-1\right)}^2}\right)\right).
\end{align*} 

Then, similar to \eqref{forecast_excess_gross_return}, we have the following expression for the excess gross forecast return  
\begin{equation} 
	\label{predictable_variation_NIG} 
	\begin{array}{llll}
		{\mathbb{E}}_tlnR^{\left(S\right)}_{t+1}-{\mathbb{E}}_tlnR^{\left(B\right)}_{t+1}
		&=\left(\frac{{\delta }^{\left(S\right)}_t{\beta }^{\left(S\right)}_t}{\sqrt{{{\alpha }^{\left(S\right)}_t}^2-{{\beta }^{\left(S\right)}_t}^2\ }}\right)
		-\left({\delta }^{\left(S\right)}_t\left(\sqrt{{{\alpha }^{\left(S\right)}_t}^2-{{\beta }^{\left(S\right)}_t}^2\ }-\sqrt{{{\alpha }^{\left(S\right)}_t}^2-{\left({\beta }^{\left(S\right)}_t-1\right)}^2}\right)\right)\\ 
		&-\left(\frac{{\delta }^{\left(B\right)}_t{\beta }^{\left(B\right)}_t}{\sqrt{{{\alpha }^{\left(B\right)}_t}^2-{{\beta }^{\left(B\right)}_t}^2\ }}\right)+\left({\delta }^{\left(B\right)}_t\left(\sqrt{{{\alpha }^{\left(B\right)}_t}^2-{{\beta }^{\left(B\right)}_t}^2\ }-\sqrt{{{\alpha }^{\left(B\right)}_t}^2-{\left({\beta }^{\left(B\right)}_t-1\right)}^2}\right)\right).  
	\end{array}
\end{equation} 
Taking the unconditional variance in \eqref{predictable_variation_NIG}, leads to the following extension of the  variation in excess returns:

\begin{equation} \label{NIG_predictable_variation} 
	\begin{array}{ccc}
		Var\left({\mathbb{E}}_tlnR^{\left(S\right)}_{t+1}-{\mathbb{E}}_tlnR^{\left(B\right)}_{t+1}\right)=\\
		Var\left( \begin{array}{c}
			\frac{{\delta }^{\left(S\right)}_t{\beta }^{\left(S\right)}_t}{\sqrt{{{\alpha }^{\left(S\right)}_t}^2-{{\beta }^{\left(S\right)}_t}^2\ }}-{\delta }^{\left(S\right)}_t\left(\sqrt{{{\alpha }^{\left(S\right)}_t}^2-{{\beta }^{\left(S\right)}_t}^2\ }-\sqrt{{{\alpha }^{\left(S\right)}_t}^2-{\left({\beta }^{\left(S\right)}_t-1\right)}^2}\right) \\ 
			-\frac{{\delta }^{\left(B\right)}_t{\beta }^{\left(B\right)}_t}{\sqrt{{{\alpha }^{\left(B\right)}_t}^2-{{\beta }^{\left(B\right)}_t}^2\ }}+{\delta }^{\left(B\right)}_t\left(\sqrt{{{\alpha }^{\left(B\right)}_t}^2-{{\beta }^{\left(B\right)}_t}^2\ }-\sqrt{{{\alpha }^{\left(B\right)}_t}^2-{\left({\beta }^{\left(B\right)}_t-1\right)}^2}\right) \end{array}
		\right) 
	\end{array}
\end{equation} 
When ${\beta }^{\left(S\right)}_t={\beta }^{\left(B\right)}_t=0,\ {\delta }^{\left(S\right)}_t={{\sigma }^{\left(S\right)}_t}^2{\alpha }^{\left(S\right)}_t,{\delta }^{\left(B\right)}_t={{\sigma }^{\left(B\right)}_t}^2{\alpha }^{\left(B\right)}_t$, and if ${\alpha }^{\left(S\right)}_t\uparrow \infty$ and ${\alpha }^{\left(B\right)}_t\uparrow \infty$, then \eqref{NIG_predictable_variation} leads to \eqref{predictable_variation_Normal}.  

However, in general, we can adjust \citet[p.~6]{LeRoy:2016}'s claim as follows:
 
\hfill\begin{minipage}{\textwidth-.5cm}
\textbf{Revised Efficiency Claim:} \textit{``The left-hand side of \eqref{NIG_predictable_variation} gives a measure of the  variation in excess returns. Eq.\eqref{NIG_predictable_variation}  shows that if the tail indices ${\alpha}^{\left(S\right)}_t$, ${\alpha}^{\left(B\right)}_t$, the asymmetric parameters ${\beta}^{\left(S\right)}_t$, ${\beta}^{\left(B\right)}_t$ and the scale parameters ${\delta }^{\left(S\right)}_t$, $\delta ^{\left(B\right)}_t$ of $z^{\left(S\right)}_{t+1/{\mathcal{F}}^{\ }_{t\ }}$ and $z^{\left(B\right)}_{t+1/{\mathcal{F}}^{}_{t\ }}$ are constant across date-$t$ events (although generally not equal to zero), then excess returns on stock are unpredictable. In that case, markets are efficient. If, on the other hand, those parameters differ according to the event, more precisely, the right-hand side of \eqref{predictable_variation_NIG} is time dependent, then a strictly positive fraction of excess returns are forecastable, so markets are inefficient.''}
\end{minipage}

\section{Numerical example}

In this section, we investigate the \textit{LL-Efficiency Claim} and \textit{Revised Efficiency Claim} using a numerical example. To determine market efficiency, we estimate  the excess return variation by applying the three models proposed in this paper.   

\subsection{Estimating the excess return variation measures using the normal distribution}
Let $R^{(S)}_{t+1})$ and  $R^{(B)}_{t+1})$ be the cumulative (gross) daily closing  stock and bond returns, respectively. Under the LLM, it is assumed that conditional on  ${\mathcal{F}}^{}_{t}$, the composite variables $z^{\left(S\right)}_{t+1\slash {\mathcal{F}}^{\ }_{t\ }}$  and  $z^{\left(B\right)}_{t++1\slash {\mathcal{F}}^{\ }_{t\ }}$ are log-normally distributed. Then, $z^{\left(S\right)}_{t+1\slash {\mathcal{F}}^{\ }_{t\ }}\sim ln\mathcal{N}\left({\mu }^{\left(S\right)}_t,{{\sigma }^{\left(S\right)}_t}^2\right)$ and  $z^{\left(B\right)}_{t+1\slash {\mathcal{F}}^{\ }_{t\ }}\sim ln\mathcal{N}\left({\mu }^{\left(B\right)}_t,{{\sigma }^{\left(B\right)}_t}^2\right)$, which leads to:

\begin{enumerate}
	\item Stock log-return, $ln R^{(S)}_{t+1}$, condition on  ${\mathcal{F}}^{}_{t}$, at $t+1$ is $\mathcal{N}\left(r_{t+1\slash {\mathcal{F}}^{\ }_{t\ }}-\frac{1}{2}\ {{\sigma }^{\left(S\right)}_t}^2,{{\sigma }^{\left(S\right)}_t}^2\right)$, where $r_t$ $t\ge 0$ is the dynamics of the risk-free rate and
	\item Bond log-return, $ln R^{(B)}_{t+1}$,  condition on  ${\mathcal{F}}^{}_{t}$, at $t+1$ is $\mathcal{N}\left(r_{t+1\slash {\mathcal{F}}^{\ }_{t\ }}-\frac{1}{2}\ {{\sigma }^{\left(B\right)}_t}^2,{{\sigma }^{\left(B\right)}_t}^2\right)$.
\end{enumerate}

To estimate the excess return variability measure, we use market indices for the
quadruple  $\left(S_t,\, \sigma_t^{S},\, r_t,\, \sigma_t^{B}  \right)$ $t\geq0$ where (i) $S_t$ $t\geq0$, the risky asset, is the SPDR S\&P 500 index;\footnote{ \url{https://www.marketwatch.com/investing/index/spx}} (ii) $\sigma_t^{S}$ $t\geq0$ is the cumulative VIX \footnote{\url{http://www.cboe.com/products/vix-index-volatility/vix-options-and-futures/vix-index/vix-historical-data}} (i.e., $\sigma_t^{S}$ represent the cumulative of VIX in $[0,t]$), (iii) $r_t$ $t\geq0$, the riskless asset as proxied by the 10-year Treasury yield \footnote{\url{https://ycharts.com/indicators/10\_year\_treasury\_rate}}, and (iv) $\sigma_t^{B}$ is the cumulative TYVIX (i.e., $\sigma_t^{B}$ represent the cumulative of CBOE 10-year Treasury Note Volatility Index (TYVIX) index in $[0,t]$).\footnote{\url{http://www.cboe.com/products/vix-index-volatility/volatility-on-interest-rates/cboe-cbot-10-year-u-s-treasury-note-volatility-index-tyvix}} The database to estimate the excess returns variation covers the period from January 2014 to December 2018 (1,258 daily observations) collected from \textit{Bloomberg Financial Markets}.  

From \eqref{predictable_variation_Normal}, the estimates for $Var\left({\mathbb{E}}_TlnR^{\left(S\right)}_{T+1}-{\mathbb{E}}_TlnR^{\left(B\right)}_{T+1}\right)$ using a fixed windows of size 252 each (one-year of daily data) are obtained by
\begin{equation}
	\label{estimate_Normal_var}
	V_{T,s}=\frac{1}{T-1}\sum^{T-1+s}_{u=s}{{\left({\gamma }_u-\ {\overline{\gamma }}_{T,s}\right)}^2}\,,\,T=252\,,\,\, s=1,\dots ,S=1006\,,
\end{equation}
where ${\gamma }_u\coloneqq -\frac{1}{2*365}{\left(\frac{\sigma_u^{S}}{100}\right)}^{2}+ \frac{1}{2*365}{\left(\frac{\sigma_u^{B}}{100}\right)}^2,\ {\overline{\gamma }}_{T,s}\coloneqq \frac{1}{T}\sum^{T-1+s}_{u=s}{{\gamma }_u}$.

Figure \ref{figure_Var_Normal} displays estimates for the variation in excess returns when the composite variables are normally distributed. The estimated values vary in the interval $(3.48, 22)$, indicating that the conditional variances of the composite variables are not constant across date-$t$ event. As to the \textit{LL-Efficiency Claim}, it is concluded that a strictly positive fraction of excess returns is forecastable, so markets are inefficient. We show that there is a problem in fitting the log-normal distribution that is reflected by the variability of the conditional variances of the composite variables. Thus, we re-estimate the  excess returns variation by fitting the NIG to the log-composite variables. 

\subsection{Estimating the excess return variation measures using the NIG distribution}

We believe that a distribution with tails heavier than the 
normal distribution can explain why the unconditional variances of the composite variables are constant across time. Here it is assumed that conditional on  ${\mathcal{F}}^{\ }_{t\ }$, the composite variables $z^{\left(S\right)}_{t+1/{\mathcal{F}}^{\ }_{t\ }}$  and  $z^{\left(B\right)}_{t+1/{\mathcal{F}}^{\ }_{t\ }}$ are   log-NIG distributed. In the other words,  $ln z^{\left(S\right)}_{t+1/{\mathcal{F}}^{}_{t}}\sim NIG\left({\mu}^{\left(S\right)}_t,{\alpha}^{\left(S\right)}_t,{\beta }^{\left(S\right)}_t,{\delta }^{\left(S\right)}_t\right)$ and
$ln z^{\left(B\right)}_{t+1/{\mathcal{F}}^{}_{t}} \sim NIG\left({\mu }^{\left(B\right)}_t,{\alpha }^{\left(B\right)}_t,{\beta }^{\left(B\right)}_t,{\delta }^{\left(B\right)}_t\right).$ \\
This leads to $\,ln R^{\left(S\right)}_{t+1/{\mathcal{F}}^{}_{t}}\sim NIG\left(m^{\left(S\right)}_t,{\alpha }^{\left(S\right)}_t,{\beta }^{\left(S\right)}_t,{\delta }^{\left(S\right)}_t\right)$ 
and $\,ln R^{\left(B\right)}_{t+1/{\mathcal{F}}^{}_{t}}\sim NIG\left(m^{\left(B\right)}_t,{\alpha }^{\left(B\right)}_t,{\beta }^{\left(B\right)}_t,{\delta }^{\left(B\right)}_t\right)$, 
where $m^{\left(S\right)}_t\coloneqq r_{t+{1}/{{\mathcal{F}}^{\ }_{t\ }}}-{\delta }^{\left(S\right)}_t\left(\sqrt{{{\alpha }^{\left(S\right)}_t}^2-{{\beta }^{\left(S\right)}_t}^2\ }-\sqrt{{{\alpha }^{\left(S\right)}_t}^2-{\left({\beta }^{\left(S\right)}_t-1\right)}^2}\right)\ \ $and  $m^{\left(B\right)}_t\coloneqq r_{t+1\slash {\mathcal{F}}^{\ }_{t\ }}-{\delta }^{\left(B\right)}_t\left(\sqrt{{{\alpha }^{\left(B\right)}_t}^2-{{\beta }^{\left(B\right)}_t}^2\ }-\sqrt{{{\alpha }^{\left(B\right)}_t}^2-{\left({\beta }^{\left(B\right)}_t-1\right)}^2}\right)$ (see the Appendix). 

Thus,  having daily return data for the S\&P500, for $\left(t,t+1,\dots,T=252\right)$, one-year of daily data, we fit the NIG distribution of daily S\&P 500 returns (assuming that these returns are iid NIG)  and find estimates for $m^{\left(S\right)}_t,\ {\alpha }^{\left(S\right)}_t,{\beta }^{\left(S\right)}_t,{\delta }^{\left(S\right)}_t.$ Moving the estimation window $\left(t+1,t+2,\dots ,T+1\right) $, we estimate $m^{\left(S\right)}_{t+1},\ {\alpha }^{\left(S\right)}_{t+1},{\beta }^{\left(S\right)}_{t+1},{\delta }^{\left(S\right)}_{t+1},$ and we continue in this manner for four years of daily data. Similarly, having daily return data for the yield on 10-year Treasury notes for  $\left(t,t+1,\dots ,T=252\right)$ we fit the NIG distribution to the daily yield of 10-year Treasury note returns (assuming that the daily returns are iid NIG)  and we find estimates for $m^{\left(B\right)}_t,\ {\alpha }^{\left(B\right)}_t,{\beta }^{\left(B\right)}_t,{\delta }^{\left(B\right)}_t.$ Moving the estimation window $\left(t+1,t+2,\dots ,T+1\right)$, we estimate $m^{\left(B\right)}_{t+1},\ {\alpha }^{\left(B\right)}_{t+1},{\beta }^{\left(B\right)}_{t+1},{\delta }^{\left(B\right)}_{t+1},$ and continue in this manner for four years of daily data.

\begin{figure}[htb]
	\centering
	\includegraphics[width=.80\textwidth]{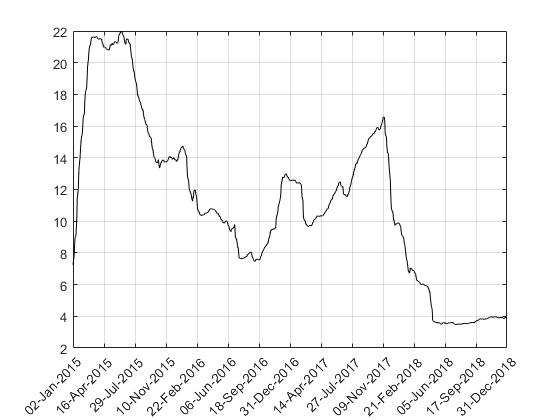}
	\caption{Estimates of the  excess returns variation when the log-composite variables are normally distributed.}
	\label{figure_Var_Normal}
\end{figure}

From \eqref{NIG_predictable_variation}, having computed (estimated) ${\alpha }^{\left(S\right)}_t,{\beta }^{\left(S\right)}_t,{\delta }^{\left(S\right)}_t,{\alpha }^{\left(B\right)}_t,{\beta }^{\left(B\right)}_t,{\delta }^{\left(B\right)}_t,t=1,\dots ,T,\ $we compute
\begin{equation}
	\label{estimate_NIG_var}
	\begin{split}
		{\chi }_t&=\frac{{\delta }^{\left(S\right)}_t{\beta }^{\left(S\right)}_t}{\sqrt{{{\alpha }^{\left(S\right)}_t}^2-{{\beta }^{\left(S\right)}_t}^2\ }}-{\delta }^{\left(S\right)}_t\left(\sqrt{{{\alpha }^{\left(S\right)}_t}^2-{{\beta }^{\left(S\right)}_t}^2\ }-\sqrt{{{\alpha }^{\left(S\right)}_t}^2-{\left({\beta }^{\left(S\right)}_t-1\right)}^2}\right)\\ 
		&-\frac{{\delta }^{\left(B\right)}_t{\beta }^{\left(B\right)}_t}{\sqrt{{{\alpha }^{\left(B\right)}_t}^2-{{\beta }^{\left(B\right)}_t}^2\ }}+{\delta }^{\left(B\right)}_t\left(\sqrt{{{\alpha }^{\left(B\right)}_t}^2-{{\beta }^{\left(B\right)}_t}^2\ }-\sqrt{{{\alpha }^{\left(B\right)}_t}^2-{\left({\beta }^{\left(B\right)}_t-1\right)}^2}\right).
	\end{split}
\end{equation}

From \eqref{predictable_variation_NIG}, the estimates for $Var\left({\mathbb{E}}_TlnR^{\left(S\right)}_{T+1}-{\mathbb{E}}_TlnR^{\left(B\right)}_{T+1}\right)$ using of fixed windows of size 252 each (one-year of daily data) are 
\begin{equation}
	\label{varaiation_estimat}
	{W}_{T,s} {=}\frac{ {1}}{ {T-1}}\sum^{ {T-1} {+} {s}}_{ {t} {=} {s}  }{{\left({\chi }_t {-} {\ }{\overline{\chi }}_{T,s}\right)}^{ {2}}}\,,\,T=252\,,\,\, s=1,\dots ,S=1006\,,
\end{equation}
where ${\overline{\chi }}_{T,s}\coloneqq \frac{1}{T}\sum^{T-1+s}_{t=s}{{\chi }_t}$.

\begin{figure}[htb]
	\centering
	\includegraphics[width=.80\textwidth]{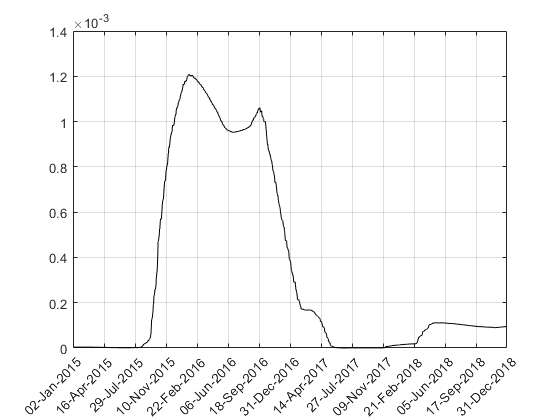}
	\caption{Estimates of the predictable excess return variation when the log-composite variables are NIG distributed.}
	\label{figure_Var_NIG}
\end{figure}

Figure \ref{figure_Var_NIG} shows the estimates of the  excess return variation across time.  Comparing Figure \ref{figure_Var_NIG} to Figure \ref{figure_Var_Normal}, one can graphically check the difference between the two methods for estimating the excess return variation. The estimates of the excess return variation in the interval $(0,0.0012)$ are constant and significantly lower when the NIG distribution is fitted compared to when the log-normal distribution is fitted.  Thus the tail indices ${\alpha}^{\left(S\right)}_t$, ${\alpha}^{\left(B\right)}_t$, the asymmetric parameters ${\beta}^{\left(S\right)}_t$, ${\beta}^{\left(B\right)}_t$ and the scale parameters ${\delta }^{\left(S\right)}_t$, $\delta ^{\left(B\right)}_t$ of the composite variables  are constant across date-$t$ events. Therefore, according to the \textit{Revised Efficiency Claim}, it is demonstrated that the excess returns for the stock market are unpredictable, and therefore markets are efficient.

The estimates of the excess return variation plotted in Figure~\ref{figure_Var_NIG} indicate that the variation of gross excess returns is attributable to poorly fitting the tail of the return distribution and that the puzzle disappears by employing a more appropriate distribution for the return data. However, to demonstrate  that the inefficiency of the market, attributable to the volatility of excess returns across time, is caused by fitting the wrong distribution, we reexamine the puzzle by fitting a new fat-tail distribution. We modify the probability distribution of the composite variables with a new heavy tail distributions and re-estimate the excess return variation to  explain the excess volatility in the market.

\subsection{Estimating the excess return variation measures using the normal compound inverse Gaussian distribution}

In this subsection, we use a new type of L\'{e}vy process relative to the NIG distribution described by \cite{Shirvani2019a}, which they call the \textit{normal compound inverse Gaussian} (NCIG) distribution. We use this distribution to  estimate the excess variation measure. The NCIG is a mixture of the normal and doubly compound of the inverse Gaussian (IG) distribution. It is a  heavy-tailed distribution with tails that are heavier compared to the NIG distribution. The moment-generating function (MGF) of the NCIG distribution has an exponential form, and it gives an explicit formula for the excess return variation measure. It seems that the NCIG distribution is an efficient distribution for composite variables because  it is a heavy tail distribution, and its MGF has an exponential form. To define the NCIG distribution, we first describe the Doubly Subordinated IG Process.\footnote{See \cite{Shirvani2019b}.} 

\noindent \textit{Definition 1: Doubly Subordinated IG Process:}
Let $T(t)$ and $U(t)$, be independent IG L\'{e}vy subordinators,\footnote{A L\textrm{\'{e}}vy subordinator is a L\textrm{\'{e}}vy process with an increasing sample path \citep[see][]{Sato:1999}.} $ T(1)\sim\,IG\left( \alpha_T,\beta_T\right)$,  $U(1)\sim\,IG\left( \alpha_U,\beta_U\right) $, then the  compound subordinator $V(t)=T\left( U(t)\right)$ has density function given by

\begin{equation}
	\label{Doubly_Subordinated}
	f_{V(1)}(x)=\frac{1}{2\pi}\sqrt{\frac{\alpha_T\alpha_U}{x^3}}\int_{0}^{\infty}u^{-\frac{3}{2}}\exp\left(-\frac{\alpha_T\left(x-\beta_T u \right)^2 }{2\beta_{T}^2 x}-\frac{\alpha_U\left(u-\beta_{U} \right)^2 }{2\beta_{U}^2 u} \right)du, \,\alpha>0, \beta>0,
\end{equation}
where $x>0$. The MGF of $V(1)$ is
\begin{equation}
	\label{MGF_IG}
	M_{V(1)}(v)=\,\exp\left( \frac{\alpha_U}{\beta_{U}} \left(1-\sqrt{1-2\,\frac{\beta_{U}^2\,\alpha_T}{\alpha_U\,\beta_{T}} \left(  1-\sqrt{1-\frac{2\beta_{T}^2}{\alpha_T}v}\right)}\right)\right),
\end{equation}
where $v\in\,\left( 0,\frac{\alpha_T}{2\beta_{T}^2}\left[ 1-\left( \frac{\alpha_T\,\beta_{T}}{2\beta_{U}^2\,\alpha_U}\right)^2 \right] \right)$.	

A special case of the Doubly Subordinated IG Process is when $\alpha_T=\alpha_U=\alpha$ and $\beta_{T}=\beta_{U}=\beta$. In this case, the NCIG is defined as follows. 

\noindent \textit{Definition 2: Normal compound inverse Gaussian}
Let $T(t)$ and $U(t)$, be independent IG L\'{e}vy subordinators, $ T(1)\sim\,IG\left( \alpha,\beta\right)$, $U(1)\sim\,IG\left( \alpha,\beta\right) $, and Let $V(t)=T(U(t))$ be a doubly subordinator IG process with MGF given by \eqref{MGF_IG}, and $B(t)$ is a Brownian motion L\'{e}vy process, denoted by $B^{\mu,\delta}$. Then the L\'{e}vy process $Z(t)=B^{\mu,\delta}\left(V(t) \right)$  is a NCIG L\'{e}vy process, denoted $Z(t)\sim NCIG(\mu,\alpha,\beta,	\delta)$, with MGF given by
\begin{equation}
	\label{MGF_NCIG}
	M_{Z(1)}(s)= 
	\exp\left( \frac{\alpha}{\beta} \left(1-\sqrt{1-2\,\beta \left(  1-\sqrt{1-\frac{2\beta^2}{\alpha}\left( s	\mu+\frac{\delta^2}{2}s^2\right) }\right)}\right)\right),
\end{equation}
where $s>0$. Furthermore, $2\beta\left(1-\sqrt{1-\frac{2\beta^2}{\alpha}\left(s\mu+\frac{\delta^2}{2}s^2\right)}\right)<1$, i.e. $\frac{\delta^2}{2}s^2+s	\mu-\frac{\alpha}{2\beta^3}\left( 1-\frac{1}{4\beta}\right) <0$.  By setting $s=iv$, the characteristic function
of $Z(t)$ is obtained,  and thus is omitted.
Now it is assumed that conditional on  ${\mathcal{F}}^{\ }_{t\ }$, the composite variables $z^{\left(S\right)}_{t+1/{\mathcal{F}}^{\ }_{t\ }}$  and  $z^{\left(B\right)}_{t+1/{\mathcal{F}}^{\ }_{t\ }}$ are   log-NCIG distributed. In the other words,  $ln z^{\left(S\right)}_{t+1/{\mathcal{F}}^{}_{t}}\sim NCIG\left({\mu}^{\left(S\right)}_t,{\alpha}^{\left(S\right)}_t,{\beta }^{\left(S\right)}_t,{\delta}^{\left(S\right)}_t\right)$ and
$ln z^{\left(B\right)}_{t+1/{\mathcal{F}}^{}_{t}} \sim NCIG\left({\mu }^{\left(B\right)}_t,{\alpha }^{\left(B\right)}_t,{\beta }^{\left(B\right)}_t,{\delta }^{\left(B\right)}_t\right).$ \\

Then, similarly to \eqref{predictable_variation_NIG}, we have the following expression for the excess gross forecast return

\begin{equation} 
	\label{predictable_variation_CNIG} 
	\begin{array}{llll}
		{\mathbb{E}}_tlnR^{\left(S\right)}_{t+1}-{\mathbb{E}}_tlnR^{\left(B\right)}_{t+1}
		&=\left({{\mu }^{\left(S\right)}_t  {\alpha }^{\left(S\right)^2}_t }\right)
		-\frac{\alpha_t^{(S)}}{\beta_t^{(S)}} \left(1-\sqrt{1-2\,\beta_t^{(S)} \left(  1-\sqrt{1-\frac{2\beta_t^{(S)^2}}{\alpha_t^{(S)}}\left( \mu_t^{(S)}+\frac{\delta_t^{(S)^2}}{2}\right) }\right)}\right)\\ 
		&-\left({{\mu}^{\left(B\right)}_t {\alpha}^{\left(B\right)^2}_t }\right)+\frac{\alpha_t^{(B)}}{\beta_t^{(B)}} \left(1-\sqrt{1-2\,\beta_t^{(B)} \left(  1-\sqrt{1-\frac{2\beta_t^{(B)^2}}{\alpha_t^{(S)}}\left( \mu_t^{(B)}+\frac{\delta_t^{(B)^2}}{2}\right) }\right)}\right).  
	\end{array}
\end{equation} 

Taking the unconditional variance in \eqref{predictable_variation_NIG} leads to the following extension of the  excess return variation measure:

\begin{equation} \label{NCIG_variation_measure} 
	\begin{array}{ccc}
		Var\left({\mathbb{E}}_tlnR^{\left(S\right)}_{t+1}-{\mathbb{E}}_tlnR^{\left(B\right)}_{t+1}\right)=\\
		Var\left( \begin{array}{c}
			\left({{\mu }^{\left(S\right)}_t  {\alpha }^{\left(S\right)^2}_t }\right)
			-\frac{\alpha_t^{(S)}}{\beta_t^{(S)}} \left(1-\sqrt{1-2\,\beta_t^{(S)} \left(  1-\sqrt{1-\frac{2\beta_t^{(S)^2}}{\alpha_t^{(S)}}\left( \mu_t^{(S)}+\frac{\delta_t^{(S)^2}}{2}\right) }\right)}\right)-\\ 
			\left({{\mu}^{\left(B\right)}_t {\alpha}^{\left(B\right)^2}_t }\right)+\frac{\alpha_t^{(B)}}{\beta_t^{(B)}} \left(1-\sqrt{1-2\,\beta_t^{(B)} \left(  1-\sqrt{1-\frac{2\beta_t^{(B)^2}}{\alpha_t^{(S)}}\left( \mu_t^{(B)}+\frac{\delta_t^{(B)^2}}{2}\right) }\right)}\right)    \end{array}
		\right) 
	\end{array}.
\end{equation} 

Similar to the NIG case, we fit the NCIG distribution to daily returns for the S\&P500 index and the yield on 10-year Treasury notes. We estimate model parameters by applying the empirical characteristic function method \citep[see][]{Yu:2003}. 

From \eqref{predictable_variation_CNIG} and the estimated values for ${\alpha }^{\left(S\right)}_t,{\beta }^{\left(S\right)}_t,{\delta }^{\left(S\right)}_t,{\alpha }^{\left(B\right)}_t,{\beta }^{\left(B\right)}_t,{\delta }^{\left(B\right)}_t,t=1,\dots ,T,$ we compute
\begin{equation}
	\label{estimate_NCIG_var}
	\begin{split}
		{\chi }_t&=\left({{\mu }^{\left(S\right)}_t  {\alpha }^{\left(S\right)^2}_t }\right)
		-\frac{\alpha_t^{(S)}}{\beta_t^{(S)}} \left(1-\sqrt{1-2\,\beta_t^{(S)} \left(  1-\sqrt{1-\frac{2\beta_t^{(S)^2}}{\alpha_t^{(S)}}\left( \mu_t^{(S)}+\frac{\delta_t^{(S)^2}}{2}\right) }\right)}\right)\\ 
		&-\left({{\mu}^{\left(B\right)}_t {\alpha}^{\left(B\right)^2}_t }\right)+\frac{\alpha_t^{(B)}}{\beta_t^{(B)}} \left(1-\sqrt{1-2\,\beta_t^{(B)} \left(  1-\sqrt{1-\frac{2\beta_t^{(B)^2}}{\alpha_t^{(S)}}\left( \mu_t^{(B)}+\frac{\delta_t^{(B)^2}}{2}\right) }\right)}\right).  
	\end{split}
\end{equation}

\begin{figure}[htb]
	\centering
	\includegraphics[width=.80\textwidth]{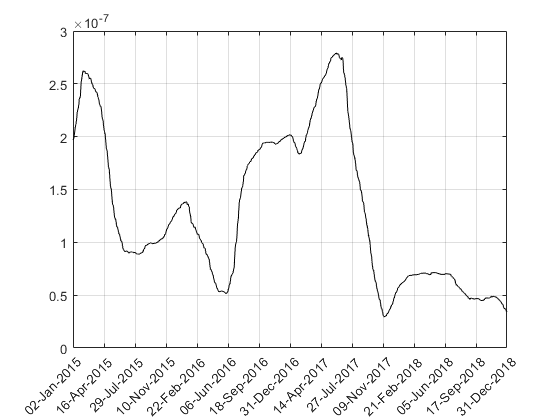}
	\caption{Estimates of the predictable excess return variation when the log-composite variables are NCIG distributed.}
	\label{figure_Var_NCIG}
\end{figure}

We compute the variation of excess returns by equation \eqref{varaiation_estimat}. Figure \ref{figure_Var_NCIG} shows the estimates of the  variation of excess returns over a rolling estimation with windows of size 252. The estimates of excess returns variation  are almost zero across date-$t$ events. These estimates are significantly lower compared to when the normal distribution and NCIG distribution are fitted.  It shows that the parameters ${\alpha}^{\left(S\right)}_t$, ${\alpha}^{\left(B\right)}_t$, ${\beta}^{\left(S\right)}_t$, ${\beta}^{\left(B\right)}_t$, ${\delta }^{\left(S\right)}_t$, and $\delta ^{\left(B\right)}_t$ are constant across date-$t$ events.
With respect to  the \textit{Revised Efficiency Claim}, it is concluded that the excess returns on the stock market are unpredictable, and that markets are efficient. 


\section{Conclusion}

\noindent 
In this study, we demonstrate that the excess volatility puzzle can be explained by fitting an appropriate distribution for the gross rate of returns for the stock market and the bond market. We empirically show that the reported market inefficiency, attributable to the volatility of excess return across time is caused by fitting the wrong distribution to historical returns. 
We fit the normal, normal inverse Gaussian, and normal compound inverse Gaussian distributions to the historical returns to evaluate the variation of excess gross returns. The results indicate that the variability of excess gross returns disappears by employing the normal inverse Gaussian and normal compound inverse Gaussian distributions to the return data. The estimated values of the excess gross returns variation are constant and markedly lower when the normal inverse Gaussian and normal compound inverse Gaussian distributions are fitted compared to when the normal is fitted. 
Thus, the excess volatility puzzle can be explained by fitting a more appropriate distribution for the excess gross return and thereby markets are efficient when the excess volatility puzzle is considered. 

Finally,we note that our approach,  fitting a proper probability distribution to capture extreme events, is not the only possible explanation for the excess volatility puzzle. However, it is a statistical model within the framework of the rational theory of finance that can be used without relying on behavioral finance assumptions when searching for an explanation of the excess volatility puzzle.


\section*{Appendix}
\noindent In this appendix, we derive the distribution of the stock gross return when the conditional distribution of the composite variable is log-NIG distributed. 

Let conditional on   ${\mathcal{F}}^{}_{t}$, $z^{\left(S\right)}_{t+1/{\mathcal{F}}^{\ }_{t\ }}=\left({\mathbb{E}}_tz^{\left(S\right)}_{t+1}\right)\left(R^{\left(S\right)}_{t+1/{\mathcal{F}}^{\ }_{t\ }}e^{-r_{t+1}/{\mathcal{F}}^{\ }_{t\ }}\right)$ be log-NIG\ distributed. This leads to  
\begin{align*}
	R^{\left(S\right)}_{t+1/{\mathcal{F}}^{\ }_{t\ }}\sim e^{r_{t+1//{\mathcal{F}}^{\ }_{t\ }}}\frac{e^{NIG\left({\mu }^{\left(S\right)}_t,{\alpha }^{\left(S\right)}_t,{\beta }^{\left(S\right)}_t,{\delta }^{\left(S\right)}_t\right)}}{\mathbb{E}e^{NIG\left({\mu }^{\left(S\right)}_t,{\alpha }^{\left(S\right)}_t,{\beta }^{\left(S\right)}_t,{\delta }^{\left(S\right)}_t\right)}}
\end{align*}
Therefore,
\begin{align*}
	lnR^{\left(S\right)}_{t+1/{\mathcal{F}}^{\ }_{t\ }}&\sim r_{t+1/{\mathcal{F}}^{\ }_{t\ }}+NIG\left({\mu }^{\left(S\right)}_t,{\alpha }^{\left(S\right)}_t,{\beta }^{\left(S\right)}_t,{\delta }^{\left(S\right)}_t\right)\\
	&-\left({\mu }^{\left(S\right)}_t+{\delta }^{\left(S\right)}_t\left(\sqrt{{{\alpha }^{\left(S\right)}_t}^2-{{\beta }^{\left(S\right)}_t}^2\ }-\sqrt{{{\alpha }^{\left(S\right)}_t}^2-{\left({\beta }^{\left(S\right)}_t-1\right)}^2}\right)\right)\\
	&\sim NIG\left(m^{\left(S\right)}_t,{\alpha }^{\left(S\right)}_t,{\beta }^{\left(S\right)}_t,{\delta }^{\left(S\right)}_t\right),
\end{align*}
where $m^{\left(S\right)}_t\coloneqq r_{t+1/{\mathcal{F}}^{\ }_{t\ }}-{\delta }^{\left(S\right)}_t\left(\sqrt{{{\alpha }^{\left(S\right)}_t}^2-{{\beta }^{\left(S\right)}_t}^2\ }-\sqrt{{{\alpha }^{\left(S\right)}_t}^2-{\left({\beta }^{\left(S\right)}_t-1\right)}^2}\right).$

\end{document}